\documentclass[aps,twocolumn,showkeys,nofootinbib]{revtex4}

\usepackage[dvips]{graphicx}
\usepackage{amssymb,amsfonts,amsmath}
\usepackage{color}
\usepackage{ulem}

\newcommand{\bi}{\begin{itemize}}
\newcommand{\ei}{\end{itemize}}

\newcommand{\ben}{\begin{eqnarray}}
\newcommand{\een}{\end{eqnarray}}
\newcommand{\benstar}{\begin{eqnarray*}}
\newcommand{\eenstar}{\end{eqnarray*}}

\newcommand{\be}{\begin{equation}}
\newcommand{\ee}{\end{equation}}

\begin{document}

\title{Unstructured intermediate states in single protein force experiments}

\author{Ivan Junier and Felix Ritort}
\email[To whom correspondence should be sent: ]{ritort@ffn.ub.es}

\affiliation{Departament de F{\'{\i}}sica Fonamental, Facultat de F{\'{\i}}sica, Universitat de Barcelona, Diagonal 647, 08028 Barcelona, Spain}

\begin{abstract}Recent single-molecule force measurements on single-domain
proteins have highlighted a three-state folding mechanism where a
stabilized intermediate state ($\mathcal{I}$) is observed on the folding
trajectory between the stretched state and the native state. Here we
investigate on-lattice protein-like heteropolymer models that lead to a
three-state mechanism and show that force experiments can be useful to
determine the structure of $\mathcal{I}$.  We have mostly found that
$\mathcal{I}$ is composed of a core stabilized by a high number of
native contacts, plus an unstructured extended chain.  The lifetime of
$\mathcal{I}$ is shown to be sensitive to modifications of the protein
that spoil the core. We then propose three types of modifications
--point mutations, cuts and circular permutations-- aiming at: 1)
confirming the presence of the core and 2) determining its location,
within one amino acid accuracy, along the polypeptide chain. We also
propose force jump protocols aiming to probe the on/off-pathway nature
of $\mathcal{I}$.
\end{abstract}


\maketitle

The recent development of single molecule experimental tools
\cite{ishijima,fisher,weiss} has allowed to investigate the fundamental
biochemical and biophysical processes occurring at a molecular level
inside the cell \cite{ritort}. For instance, the folding of proteins
\cite{finkelstein,junier,fersht,onuchic} can nowadays be studied by
manipulating {\it one} protein at a time \cite{ritort}.  Examples are the titin molecule pulled by AFM \cite{afm1,afm2} or the {\it E. coli} 
$155$-residues RNase H protein \cite{nienhaus,nienhaus2}
pulled by optical tweezers \cite{cecconi}. At low denaturant concentration, FRET
measurements have shown the presence of highly compact denaturated
states \cite{nienhaus,nienhaus2} whose existence was expected from
previous bulk experiments \cite{raschke}. The latter suggests a
hierarchical folding mechanism where the folding of the protein to the
native state is preceded by a fast collapse of the most stable region of
the native structure. The formation of a structure that has a short
lifetime and many native contacts has been observed during the folding
of many single-domain proteins \cite{finkelstein,daggett}.  On the other
hand, recent experiments using optical tweezers have investigated the
unfolding/folding transition of the RNase H protein under the action of
a mechanical force applied at the two ends of the molecule
\cite{cecconi}.  These experiments show the stabilization of an
intermediate state at forces around $5 \mbox{pN}$ \cite{cecconi}.  The
protein is observed to exist in three different states: the stretched
($\mathcal{S}$), the intermediate ($\mathcal{I}$) and the native
($\mathcal{N}$) states\footnote{Abbreviations: $\mathcal{N}$, native state; 
$\mathcal{I}$, intermediate state; $\mathcal{S}$, stretched state;
$\mathcal{E}$, early state.}.  Using thermodynamic considerations it has been
argued that $\mathcal{I}$ is identical to the early state
($\mathcal{E}$) that forms at zero force and room temperature \cite{cecconi}.  The
experimental results also suggest that $\mathcal{I}$ is an obligatory
step in the folding pathway from $\mathcal{S}$ to $\mathcal{N}$,
hereafter referred to as an intermediate {\it on-pathway}.

The determination of the structure of generic unstructured states,
i.e. that lack a well-structured three-dimensional fold, is a major
experimental challenge in modern biophysics.  A well-known example is
the molten-globule state sometimes observed in thermal denaturation in
proteins \cite{finkelstein}.  The identification of the unstructured
states is limited by their large structural fluctuations that make
usual techniques (X-ray or NMR) poorly predictive. On the other hand,
growing evidence shows that a large number of proteins are
intrinsically unstructured and contain a fair amount of disordered
regions \cite{DysWri05}. The use of new experimental techniques aiming to 
probe unstructured states is therefore a question of great interest.

Is there any connection between the intermediate states that have been detected in AFM and 
optical tweezers experiments and the intrinsically disordered states observed in many proteins? 
Is it possible to extract useful information about the structure of the intermediate state 
observed in single molecule pulling experiments by designing specific experimental protocols?.
To address such questions we use a phenomenological approach based on the numerical investigation of
on-lattice heteropolymers in the presence of mechanical force
\cite{shakhnovich}. This class of models contain the minimal number of
ingredients necessary to capture the basic phenomenology
(thermodynamics and kinetics) of the folding transition problem. In addition, they
are simple enough to allow exhaustive statistical studies that are
difficult to carry out with other more accurate and realistic
descriptions of proteins. In contrast to simple two-state models, on-lattice heteropolymers are
phenomenological models where the molecular extension that reflects
the internal configuration of the protein is the natural reaction
coordinate \cite{socci}. 

By introducing mechanical force in the analysis \cite{klimov}, we show
how it is possible to reproduce and interpret the three-state behavior
observed in the experiments.  We numerically investigate several
topologies of the native structure and find that they generally lead
to a three-state scenario in the presence of mechanical force.  The
new intermediate state ($\mathcal{I}$) is typically composed of a
compact core with a high number of native contacts, plus an
unstructured extended chain. Moreover, $\mathcal{I}$ is not
necessarily identical to the early state ($\mathcal{E}$) that forms
when folding at zero force.

\begin{figure}[t]
\begin{center}
\centerline{\includegraphics[width=.5\textwidth]{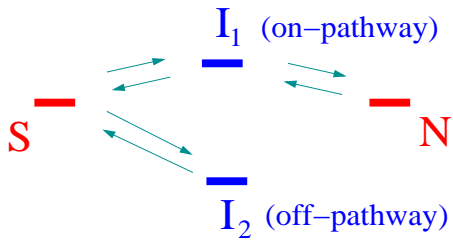}}

\vspace*{-0.5cm}

\caption{Kinetic scheme of a folding reaction with different types of
  intermediate states. $\mathcal{I}_2$ is off-pathway (misfolded)
since it is not directly connected to $\mathcal{N}$ whereas
$\mathcal{I}_1$ is on-pathway.
\label{4state}}
\end{center}
\end{figure}

We then show how the structure of the intermediate state
$\mathcal{I}$, that has been observed in single molecule pulling
experiments, can be determined by means of specific experimental
protocols that have been used in protein biophysics in different
contexts.  We propose experimental single protein force protocols
that introduce modifications in the amino acid sequence of the protein
to infer information about the structure of $\mathcal{I}$. We propose
three techniques based on i) single amino acid mutations, ii) cutting
off the polypeptide chain at various lengths and iii) circular
permutations of the protein. These techniques lead to the location of
the core due to the fact that the system $\mathcal{S} \leftrightarrow
\mathcal{I}$ undergoes a transition when the modifications involve
amino acids of the core. These protocols could be also used in
the future to unveil the local structure of globally unstructured
proteins that contain a mixture of disordered and ordered regions
\cite{DysWri05}.

Finally, by investigating the folding kinetics at different solvent conditions,
we have also found the presence of other intermediate states that, we
show, are misfolded states (Fig. \ref{4state}). In contrast with
on-pathway states, misfolded states are {\it off-pathway}: starting from
such state, the folding pathway to $\mathcal{N}$ must pass through
$\mathcal{S}$.  Although off-pathway and on-pathway states may be hard
to distinguish (e.g. when they have the same molecular extension), we
show that a force jump protocol is useful to quantify the fraction
of on/off-pathway trajectories that lead to on/off pathway
states respectively.

\begin{figure}[t]
\begin{center}
\centerline{\includegraphics*[width=.25\textwidth]{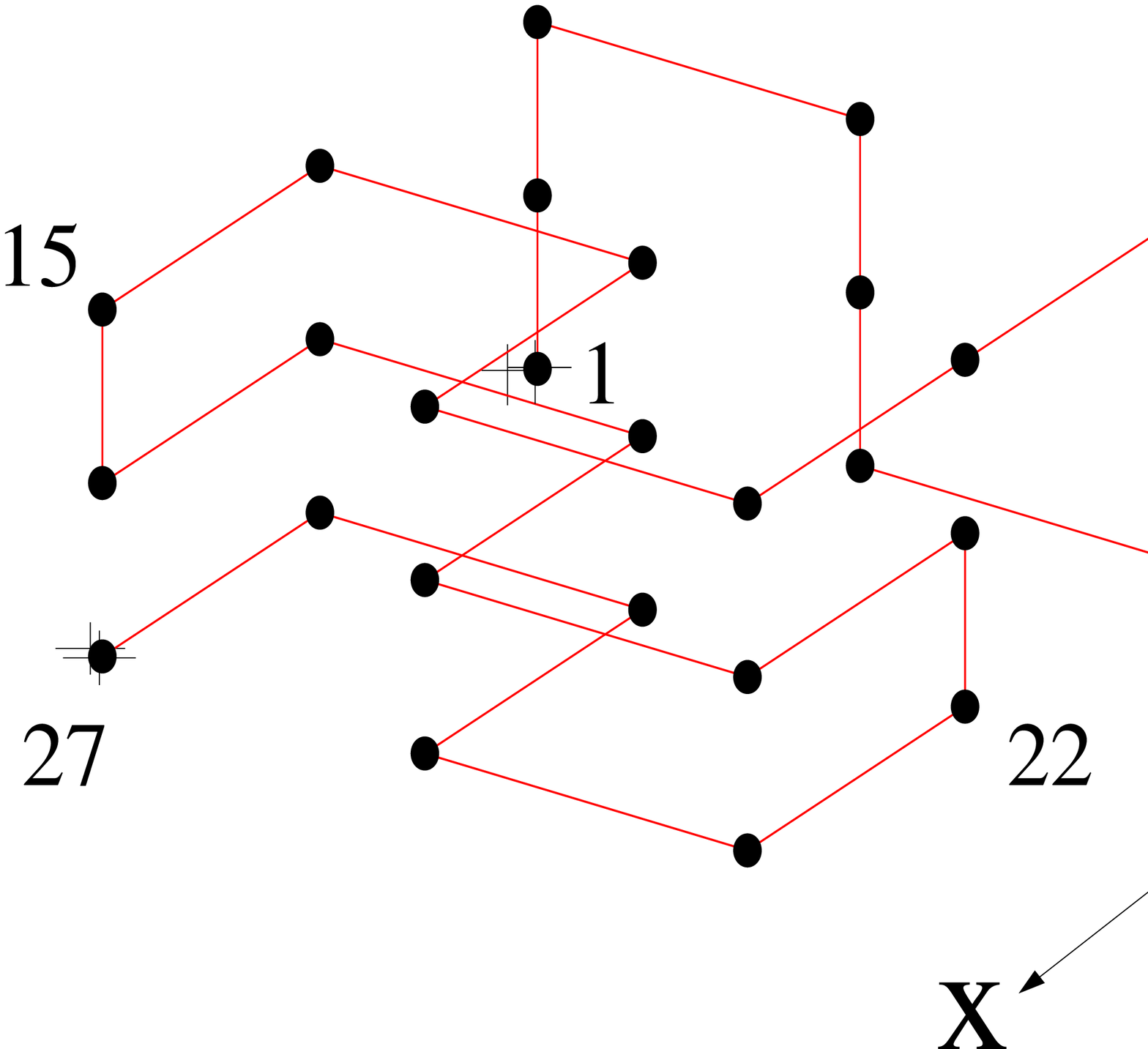}
\includegraphics[width=.25\textwidth]{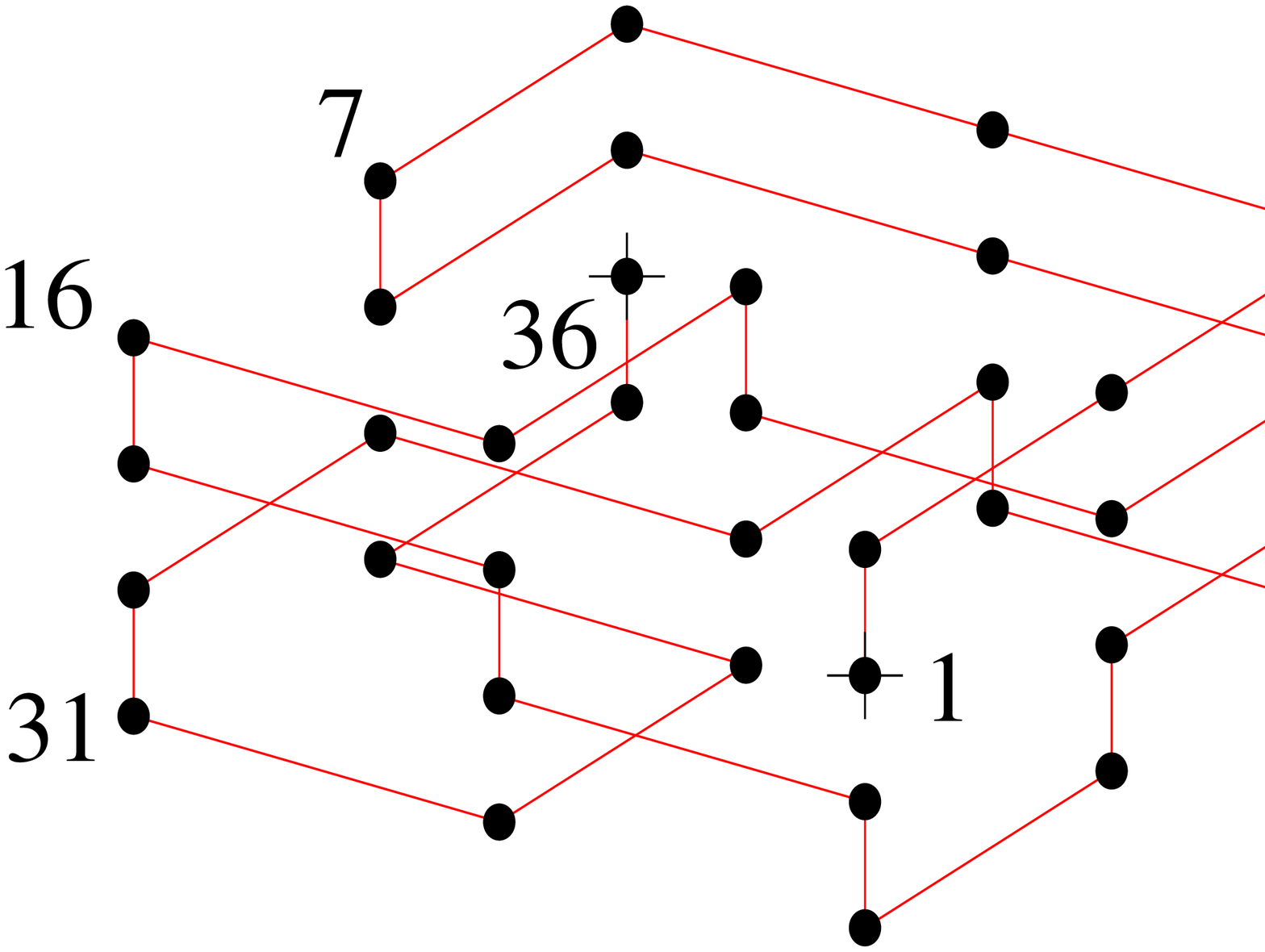}}
\caption{Two archetypal native topologies of designed heteropolymers on a cubic lattice.
Left: Structure {${\bf S_1}$}, $N=27$. Right: Structure {${\bf S_2}$}, $N=36$. The numbers indicate the positions $n$ of the monomers along the
chain. The crosses indicate the two ends of the chain.
\label{structures}}
\end{center}
\end{figure}

\section{Three-state proteins}
\label{DLH}

Following the sequence optimization procedure of Shakhnovich and Gutin
\cite{shakhnovich}, we design heteropolymers on a cubic lattice that
fold into a {\it unique} compact structure
(Fig. \ref{structures}). The heteropolymer consists of a chain of
monomers indexed by $i$ ($1 \leq i \leq N$) with nearest neighbor pair
interactions $E_{ij}$ between monomers $i,j$ that are not contiguous
along the chain. The values of $E_{ij}$, which determine the native
configuration, are obtained following an optimization algorithm
\cite{shakhnovich} starting from an initial set of interactions
$E_{ij}^0$ and a given topology of the native structure, i.e. a given
chain configuration in $\mathcal{N}$.  We note that, by definition of
the model, several sets of interactions $E_{ij}$ can be associated to
identical topologies of the native state.  The values of $E_{ij}^0$,
and hence of $E_{ij}$, are drawn from a Gaussian distribution of zero
mean and variance $\Delta^2$. $\Delta$ is measured in units of $k_B
T^*$ where $k_B$ is the Boltzmann constant and $T^*$ is a reference
temperature that we fix to $300 \mbox{K}$.  The dynamics of the
heteropolymer consists in the standard "coin and crankshaft"
Monte-Carlo dynamics with Metropolis rates \cite{hillorst} (elementary
moves are shown in the illustration). Note that these types of moves
might not be optimally suited for pulling experiments since they
transmit stress very slowly over long straight chains. However, we
still expect that the generality of our results goes beyond the
details of the local dynamics we use for the on-lattice
heteropolymers.
\begin{center}
\centerline{\includegraphics*[width=.4\textwidth]{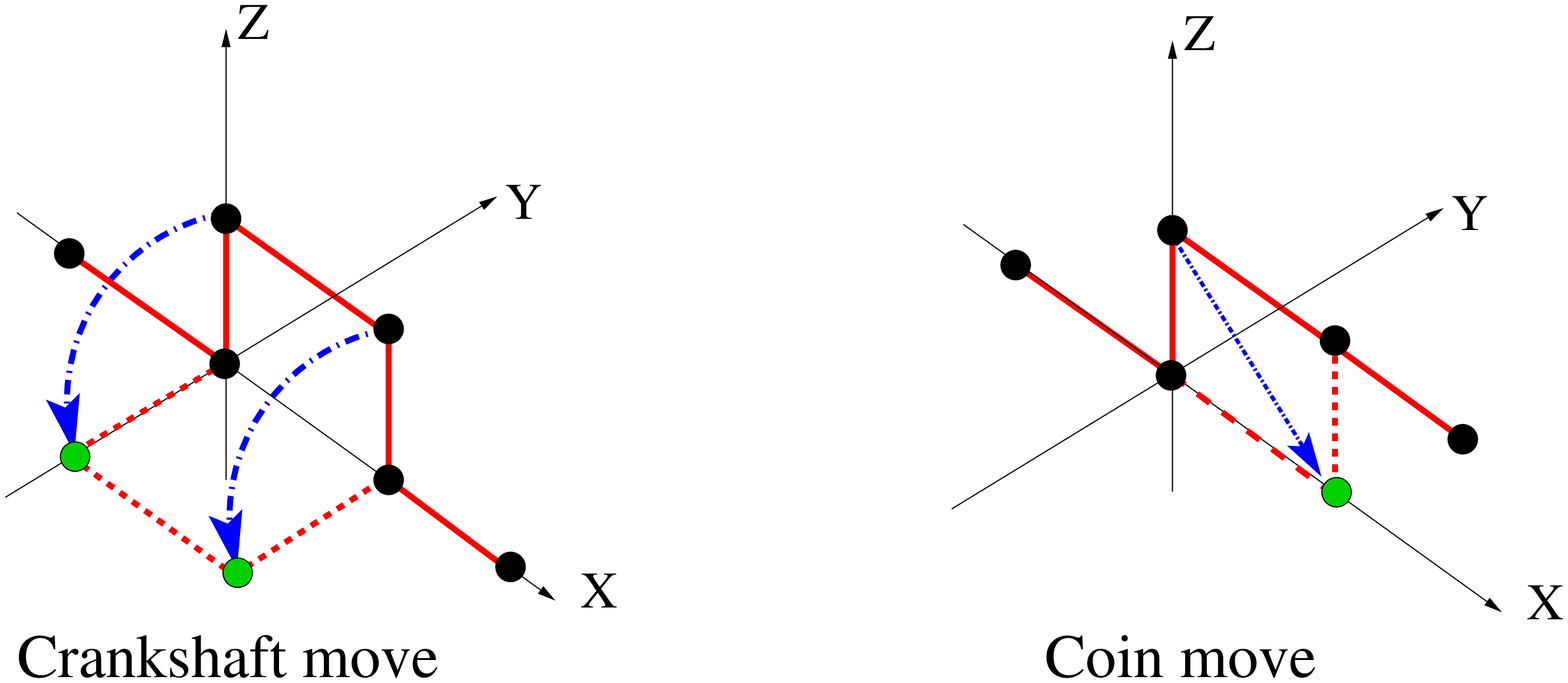}}
\end{center}

The timescale is fixed by the unit of Monte-Carlo steps
that we set to $100$ ns, a value that leads to results in quantitative
agreement with experimental results (e.g. \cite{cecconi}).  In this type
of model, the values of $E_{ij}$ correspond to specific short-range 
tertiary contacts along the protein chain.  Although long-range interactions,
side-chain interactions and other short-scale details of proteins (such
as the secondary structural motifs) are not included in the model, such
designed heteropolymers have been shown to display folding properties
that are similar to those of single-domain proteins \cite{shakhnovich}.  The
results we show here are quite general and have been reproduced with
different native structures. However, for the sake of clarity,
throughout this paper we present results for 
two archetypal compact structures ${\bf
S_1}$, ${\bf S_2}$ (Fig. \ref{structures}) whose sizes are respectively
$N=27, 36$.  These correspond to small globular proteins with a number
of residues in the range of $50-100$ \cite{onuchicw}.

\begin{figure}[t]
\begin{center}
\centerline{\includegraphics[width=.5\textwidth]{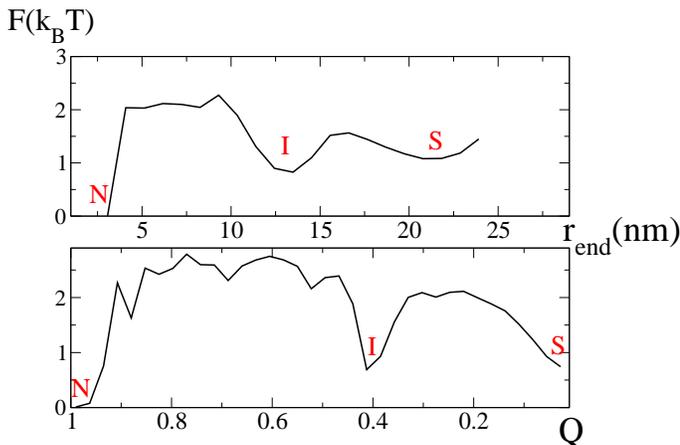}}
\caption{Free energy profiles projected along $r_{\mbox{\scriptsize
end}}$ and $Q$ for the structure ${\bf S_2}$. $\Delta = 1.2 \, k_B T^*$
and $f=9$ pN. Three main states can be defined: the native state
($\mathcal{N}$), an intermediate state ($\mathcal{I}$) and the stretched
state ($\mathcal{S}$). They correspond to the deeper local minima along
the free energy profiles.  The values for $r_{\mbox{\scriptsize end}}$ and $Q$ have
been averaged over $10 \mu s$.
\label{free}}
\end{center}
\end{figure}
  To characterize the state of the heteropolymer, we monitor the
temporal evolution of the end-to-end distance $r_{\mbox{\scriptsize
end}}$ of the molecule and the percentage of native contacts $Q$ ($0
\leq Q \leq 1$). The lattice spacing is set equal to $1$ nm for the
heteropolymer to have contour lengths that are similar to those of
proteins studied in experiments (e.g. \cite{cecconi}). A state is
defined as the location of a minimum in the free energy projected along
$Q$ or $r_{\mbox{\scriptsize end}}$ (see Fig. \ref{free}). Due to
the discrete nature of the on-lattice heteropolymer, the free energy
landscape along $r_{\mbox{\scriptsize end}}$ is a rugged surface (see
for instance \cite{socci}). Intermediate states then appear as highly
roughed basins that can be better identified by ensemble or time
averaging of the values of $r_{\mbox{\scriptsize end}}$ and $Q$ over a
finite bandwidth. In this way, we obtain smooth free energy
landscapes in space and $Q$ with well defined minima (see
Fig. \ref{free}).  Small single-domain proteins are commonly described
as two-state systems having two possible conformations: native and
denaturated \cite{finkelstein}.  In experiments, by varying the
concentration of denaturant one finds a first-order like
transition where both states coexist \cite{finkelstein} -see however
\cite{munoz} for exceptions to this general result.  In the presence of
applied mechanical force, cooperative transitions take place between the
native state $\mathcal{N}$ and a stretched state $\mathcal{S}$ as
observed in single molecule AFM measurements in engineered polyproteins
\cite{rief} and in RNA pulling experiments using optical tweezers
\cite{liphardt}. In order to introduce mechanical force in the lattice
we must avoid the lattice anisotropy effects that act as kinetic traps
for the rotational degrees of freedom \cite{klimov}. To this end, we add
a term of the type $-\vec f \cdot \vec r_{\mbox{\scriptsize end}}$ where
$\vec f$ is a force of constant modulus (measured in units of
$\Delta/a$) that is always aligned with the end-to-end vector $\vec
r_{\mbox{\scriptsize end}}$ of the heteropolymer \cite{klimov} -- see
supplementary material. We have verified that a two-state system under
the action of a (moderate) mechanical force leads to an exponential
folding/unfolding times distribution (Fig. S1 and S2 in the Supp. Mat.).
\begin{figure}[t]
\begin{center}
\centerline{\includegraphics*[width=0.3\textwidth]{fig4a.eps}
\includegraphics[width=0.22\textwidth]{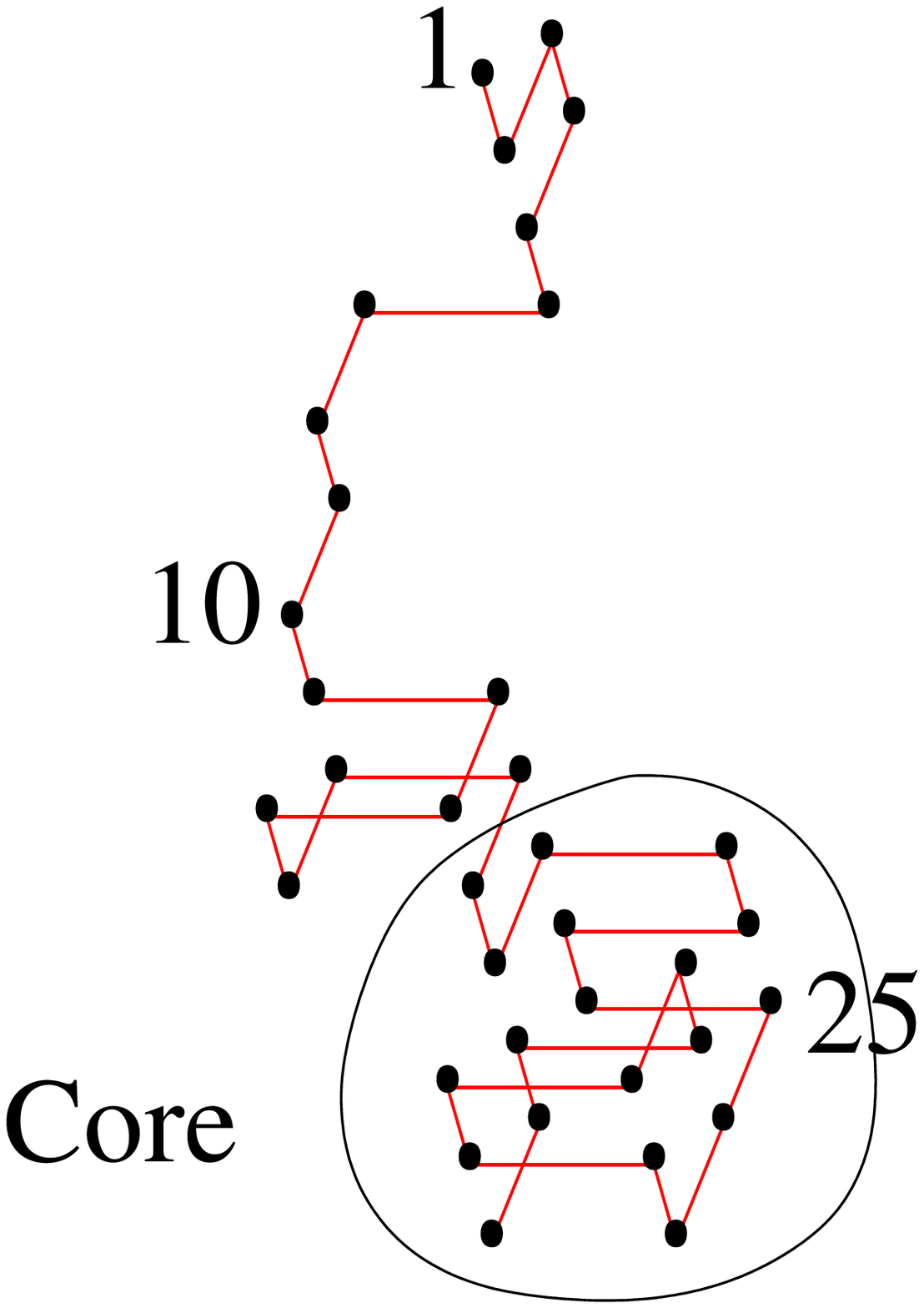}}
\centerline{\includegraphics*[width=0.3\textwidth]{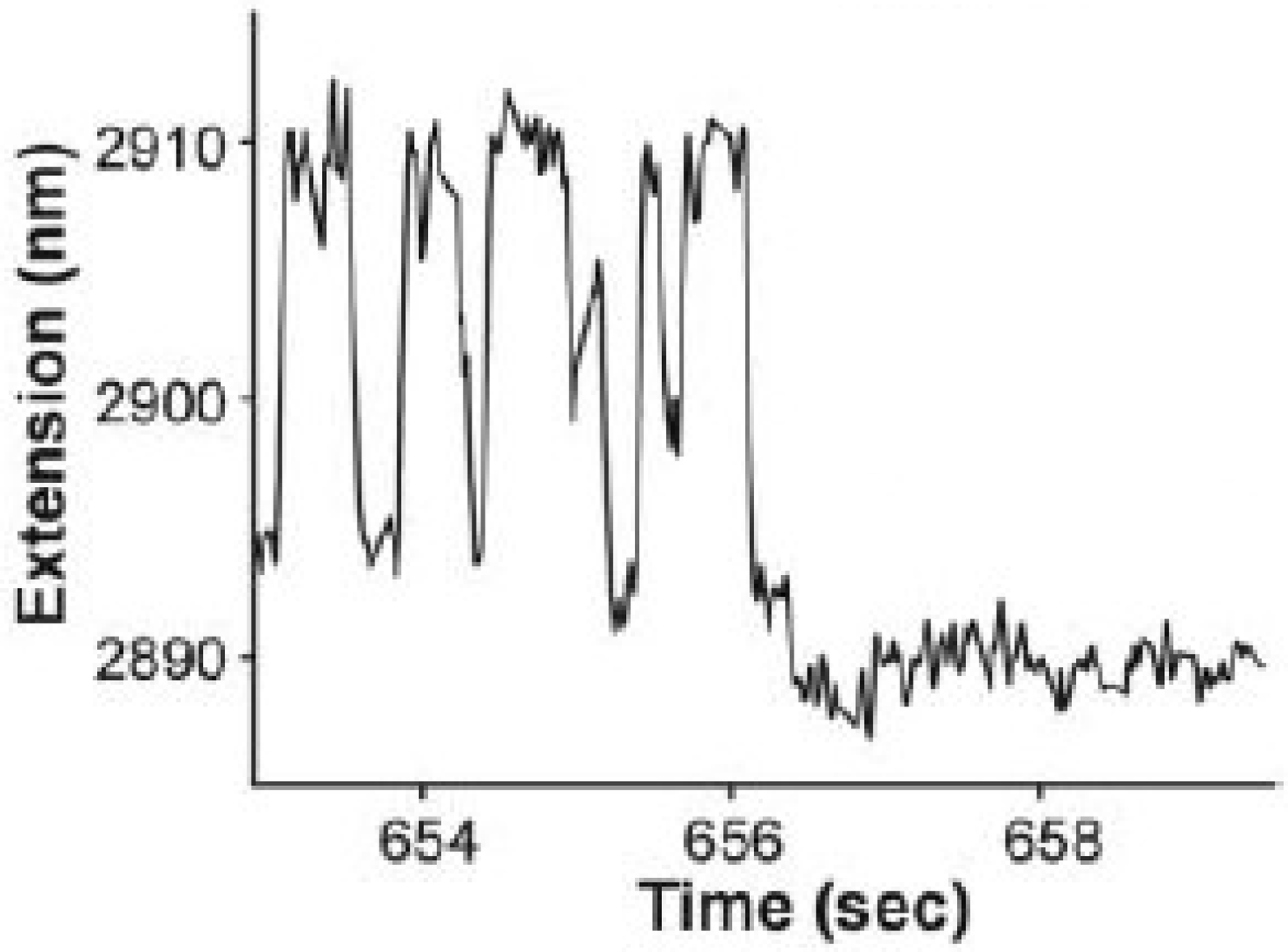}}
\caption{Upper left: Three-state behavior in $\bf S_2$. $\Delta=1.35 \,
k_B T^*$ and $f=10.1$ pN.  Upper right: Typical structure, composed of a
core plus an extended chain, of a configuration in
$\mathcal{I}$. Lower panel: Experimental trace of the RNase H protein
at constant force ($f \sim 6$ pN) using optical tweezers (taken from
\cite{cecconi}).
\label{3state}}
\end{center}
\end{figure}

\subsection*{Intermediate and misfolded states}

Starting from $\mathcal{S}$ and by further decreasing the force down
to zero, a single-domain protein shows a cooperative transition to
$\mathcal{N}$ at a given value of the force.  Our simulations show
that several structures that exhibit a two-state behavior in a given
range of temperatures, also show a three-state behavior under the
action of mechanical force at lower temperatures -- or equivalently at
larger values of $\Delta$ at the fixed temperature $T^*$. In
Fig. \ref{3state}, we show the three-state behavior by plotting the
temporal evolution of both $r_{\mbox{\scriptsize end}}$ and $Q$,
starting from a random initial configuration. The three-state
mechanism has been observed for different matrices $E_{ij}$, i.e. for
different energy values and different topologies of the native
structure.

\begin{figure}[t]
\begin{center}
\centerline{\includegraphics*[width=0.45\textwidth]{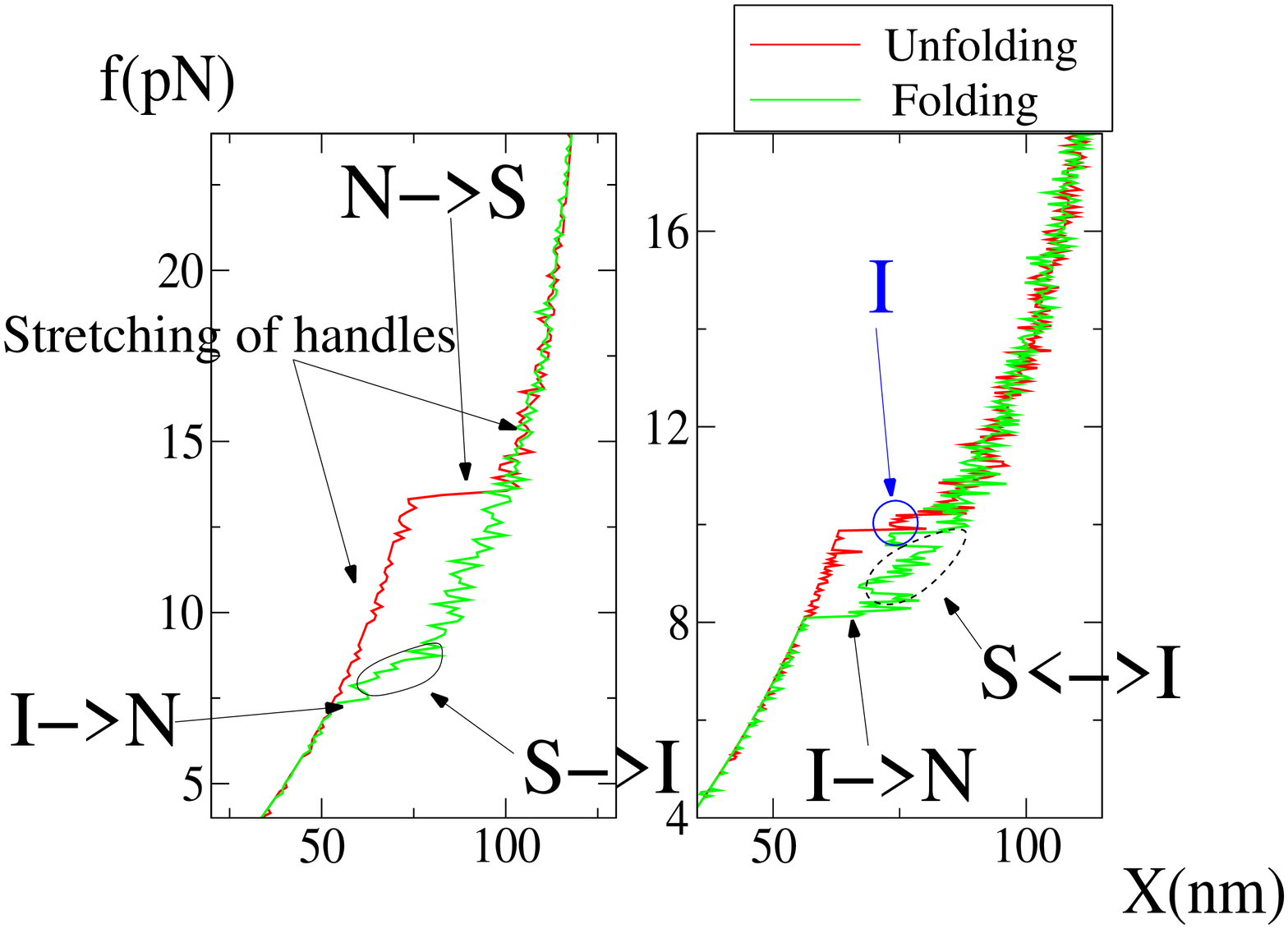}}
\caption{Force-extension curve for ${\bf S_2}$ with $\Delta = 1.2 \,
k_B T^*$ in a pulling protocol with different loading rates $r$.
Left, $r=4 \, \mbox{pN.s}^{-1}$. Right, $r= 0.4 \, \mbox{pN.s}^{-1}$.
We have included the contribution of the handles (modeled as a freely
jointed chain). The total extension (protein plus handles) is
equal to $X=r_{\mbox{\scriptsize end}}+x_{\mbox{\scriptsize FJC}}$
where $x_{\mbox{\scriptsize FJC}}=100\coth(fa/k_BT)-k_BT/fa$
corresponds to the extension of the freely jointed chain at force
$f$. At large loading rate $r$, the unfolding transition
$\mathcal{N} \to \mathcal{S}$ is of the all-or-none type
\cite{cecconi} whereas at lower loading rates (right panel), the
intermediate state $\mathcal{I}$ along the transition (blue circle)
can be resolved. At low rates, we also observe multiple transitions
between $\mathcal{S}$ and $\mathcal{I}$ during the refolding (black
dashed circle).
\label{ramp}}	
\end{center}
\end{figure}

Fig. \ref{3state} shows that the final folding stage takes place from
$\mathcal{I}$ suggesting that $\mathcal{I}$ is on-pathway. Sometimes,
however, the transition from $\mathcal{S}$ to $\mathcal{N}$ does not
go through $\mathcal{I}$ (See Fig. S3). Although this transition is
rare, it clearly shows that the folding pathway is non unique.  By
repeatedly pulling and relaxing the protein at loading rates
equivalent to those used in the experiments \cite{cecconi}, we observe
an all-or-none unfolding transition of $\mathcal{N}$
(Fig. \ref{ramp}). At much lower loading rates, we observe large
fluctuations in the molecular extension due to the presence of
$\mathcal{I}$ (Fig. \ref{ramp}), a result that is consistent with the
simulations in \cite{klimov}. These features of the force-extension curves could be checked in future single molecule experiments.

\paragraph*{{\bf The native topology.}}

\begin{figure}
\includegraphics[scale=0.35]{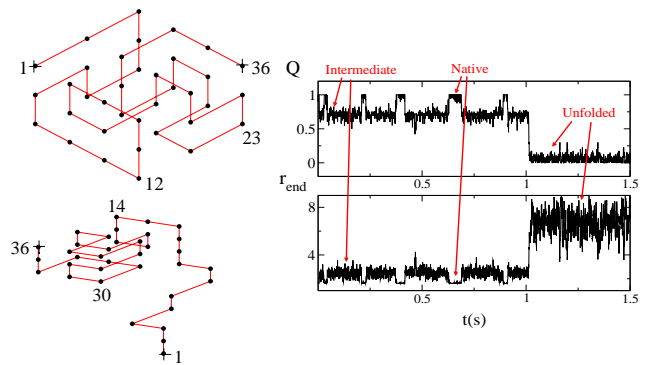}
\caption{Example of a structure for which the kinetic barrier between
$\mathcal{I}$ and $\mathcal{N}$ is smaller than that between
$\mathcal{I}$ and $\mathcal{S}$. In this case, during the unfolding
transition ($\mathcal{N} \to \mathcal{S}$), we can observe a transient
regime preceding the transition where the molecule switches between
$\mathcal{N}$ and $\mathcal{I}$. $\Delta=1.09 \, k_B T^*$ and $f=9.3$ pN. The leftmost
lower figure shows a typical configuration of the heteropolymer chain
in $\mathcal{I}$.}
\label{versatile}
\end{figure}

\begin{figure}
\includegraphics[scale=0.35]{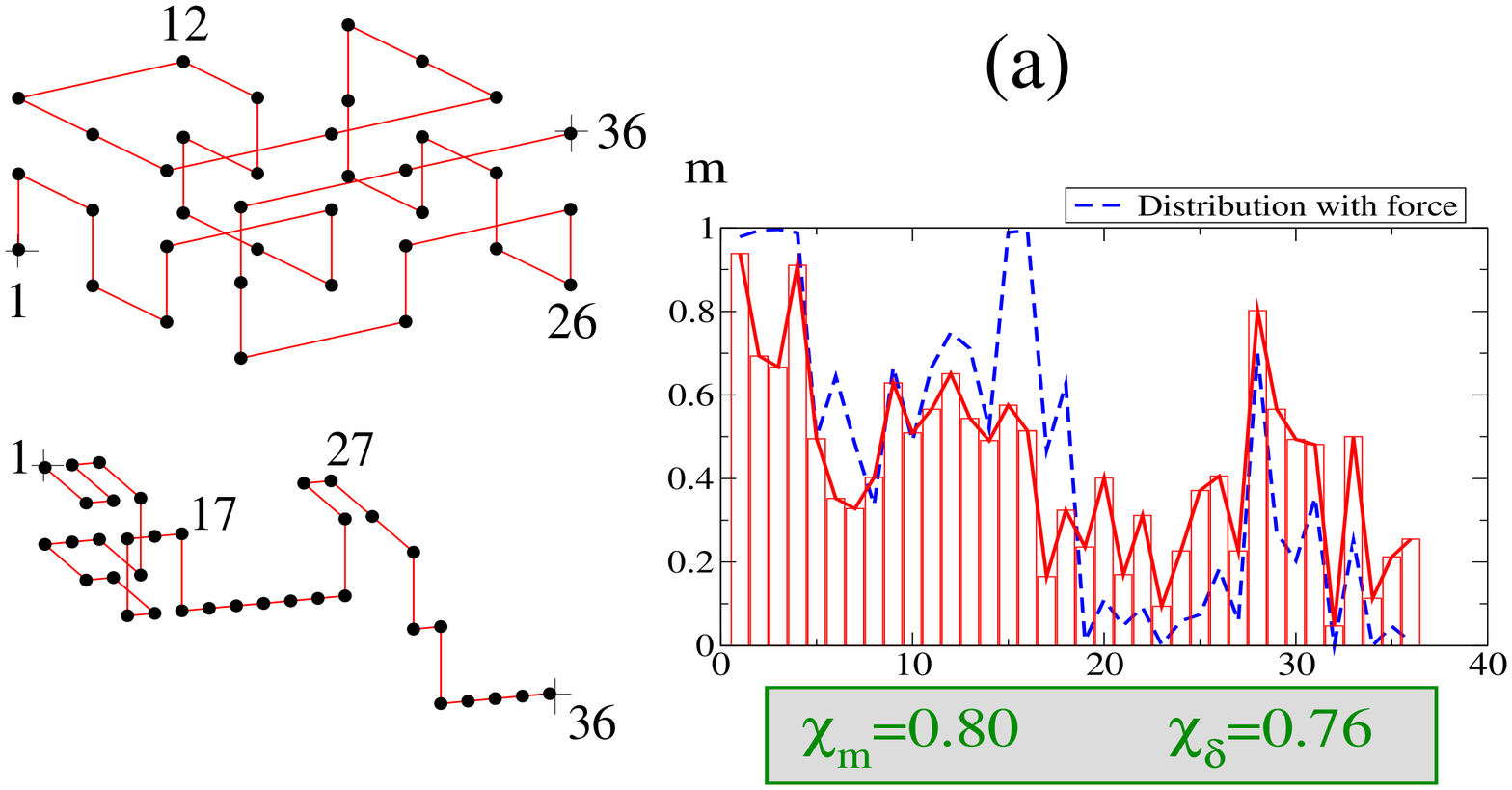}
\includegraphics[scale=0.35]{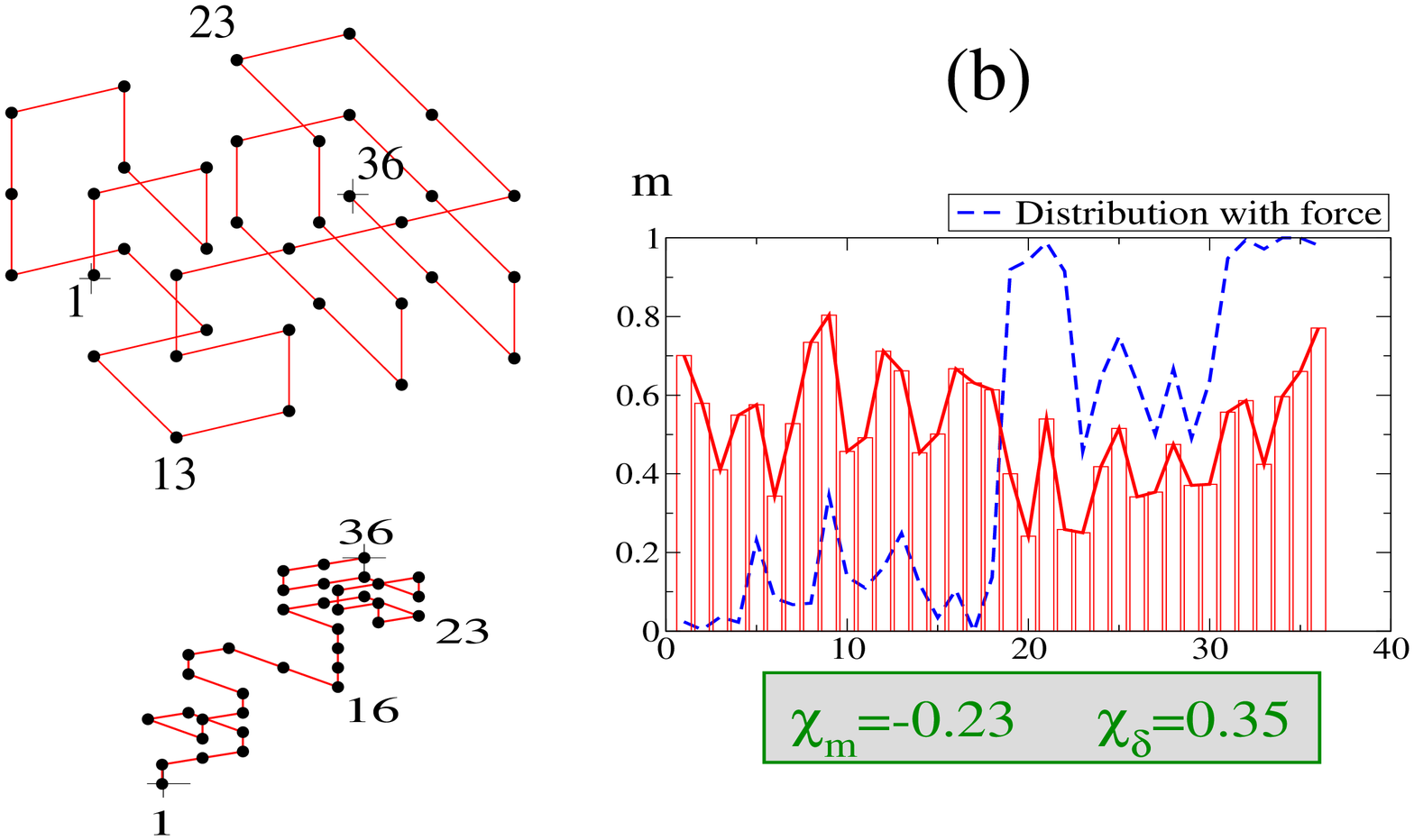}
\includegraphics[scale=0.35]{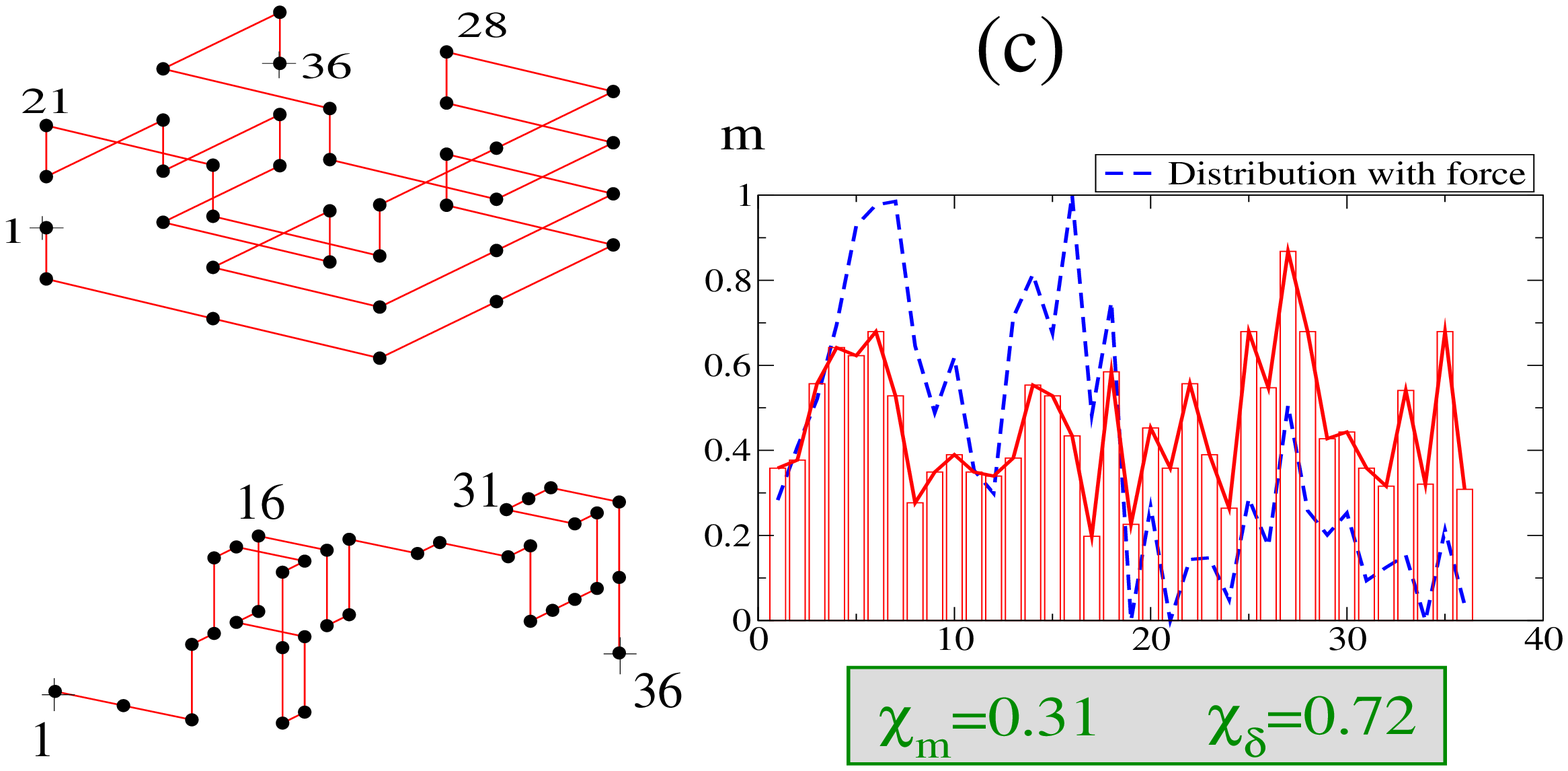}
\includegraphics[scale=0.35]{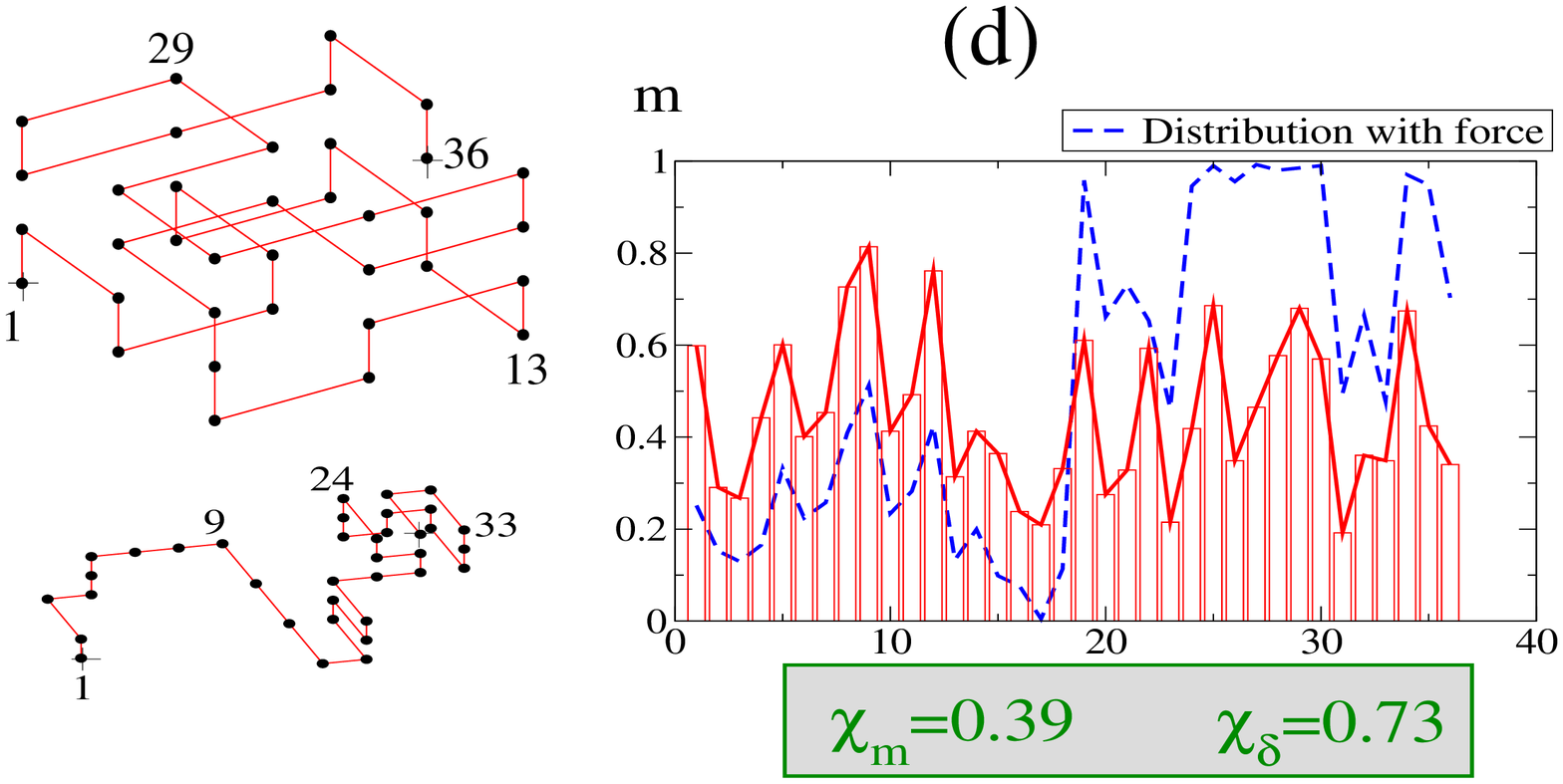}
\caption{Four examples of structures showing a three-state
behavior and having different degrees of correlation between
$\mathcal{I}$ and $\mathcal{E}$. For each figure, the upper
panel on the left shows the native structure whereas the lower
panel on the left shows a typical configuration of the heteropolymer
chain in $\mathcal{I}$. Fig \ref{examples}a: $\Delta=1.09 \, k_B T^*$
and $f=7.6$ pN. Fig
\ref{examples}b: $\Delta=  1.09 k_B T^*$ and $f=6.7$ pN. Fig \ref{examples}c:
$\Delta=1.14 k_B T^*$
and $f= 7.5$ pN.  Fig \ref{examples}d: $\Delta=1.09 k_B T^*$
and $f= 7.2$ pN.}
\label{examples}
\end{figure}

For each heteropolymer we have determined the topology of $\mathcal{I}$,
i.e. the configuration of the chain in the state $\mathcal{I}$.
Remarkably, we have always obtained a state composed of a compact core
whose contacts are mainly native plus a chain that is extended and hence
that has few native contacts --see Fig. \ref{3state}, \ref{versatile}
and \ref{examples}.

\begin{figure}
\centering\includegraphics[scale=0.25]{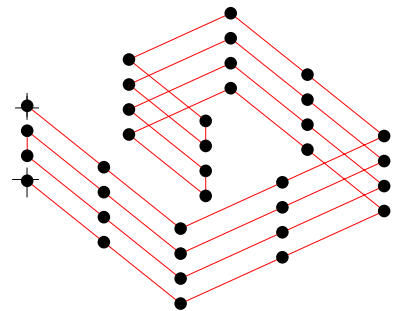}
\caption{Example of a  structure ($N=36$) for which we have not observed any three-state mechanism under any conditions. }
\label{spring}
\end{figure}

Next, we have investigated the folding mechanism for different matrices
of energies $E_{ij}$ that keep the same native topology, i.e. the same
chain configuration in $\mathcal{N}$. In most of the cases, we find a
three-state behavior where $\mathcal{I}$ shows a structure formed by the
same compact core plus an extended random coil (see also Fig. S4 in
Supp. Mat.). Because the core, and hence $\mathcal{I}$, is identical for
all cases, this suggests a strong correlation between the three-state
behavior and the topology of the native structure, independently of
the precise values of the energies $E_{ij}$.  In addition, we have
checked that some topologies never lead to the formation of an
intermediate state. As an example, the structure shown in
Fig. \ref{spring} does not lead to a three-state mechanism for any
combination of temperature and force values. However, we are not able to
give the feature list that must verify a native structure in order to
show a force induced three-state behavior in a given range of
temperatures. In contrast, as we shall see below, we have found several
different structures that show three-state behavior at sufficiently low
temperatures.

\paragraph*{\bf A versatile intermediate state}

The experiments on RNase H \cite{cecconi} and our simulations using $\bf S_2$ 
suggest that the free energy barrier separating $\mathcal{I}$ and
$\mathcal{N}$ is higher than the free energy barrier separating $\mathcal{I}$
and  $\mathcal{S}$ (Fig. \ref{free}). This explains that in some range of
force and temperature the folding transition to $\mathcal{N}$ starting from
$\mathcal{S}$ is preceded by a transient regime where the molecule 
switches between $\mathcal{S}$ and $\mathcal{I}$ --see Fig. \ref{3state}.
We have found other scenarios where the free energy barrier between
$\mathcal{N}$ and $\mathcal{I}$ is smaller than that between
$\mathcal{S}$ and $\mathcal{I}$. In this case we observe, at some
force and temperature values, a behavior symmetric to the previous one, i.e. a
switching behavior between $\mathcal{N}$ and $\mathcal{I}$ that
precedes the unfolding transition from $\mathcal{N}$ to $\mathcal{S}$ --see
Fig. \ref{versatile}.

\begin{figure}[t]
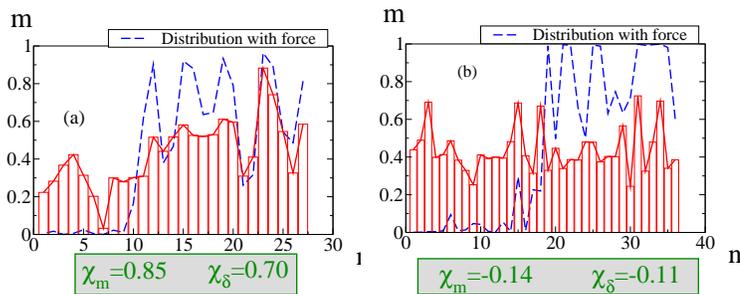

\begin{center}
\centerline{\includegraphics*[width=0.27\textwidth]{fig6a.eps}
\includegraphics[width=0.27\textwidth]{fig6b.eps}}
\caption{Average number $m$ of native contacts as a
function of the position $n$ of the monomer along the chain.  The
histograms (in red) correspond to the number of native contacts
in the early state ($\mathcal{E}$) whereas the blue dashed lines
correspond to the number of native contacts in $\mathcal{I}$ at (a)
$\Delta =1.43 \, k_B T^*$, $f=10.6$ pN for $\bf S_1$ and (b)
$\Delta=1.2\, k_B T^*$, $f=9$ pN for $\bf S_2$.
\label{compo}}
\end{center}
\end{figure}

\paragraph*{{\bf The intermediate state with and without force.}}
We have investigated whether $\mathcal{I}$ corresponds to the early
compact structure $\mathcal{E}$ that forms, starting from a random
initial configuration, during the folding at zero force and at the same
temperature. We find that sometimes both states are correlated, whereas
in other cases they are not.

At zero force, $\mathcal{I}$ is not well-defined since it is not a
local minimum along $Q$ or $r_{\mbox{\scriptsize end}}$.  As a
consequence, we have used a heuristic method to determine the state
$\mathcal{E}$ that has to be compared with $\mathcal{I}$. The
procedure is based on the fact that, in average, $Q$ monotonically
tends to $1$ during the folding transition. Therefore, for a given
random initial condition ($Q \approx 0$), during one folding
trajectory at zero force, we record the first configuration that has a
value of $Q$ identical to the value of $Q$ in $\mathcal{I}$. We then
define the state $\mathcal{E}$ as the ensemble of these first
configurations that are obtained by sampling different random initial
conditions and different noise histories.  In this ensemble of
configurations, we compute the average number $m$ of native contacts
for a given monomer as a function of its position $n$ along the chain.
The distribution $m(n)$ is then compared to that obtained for
$\mathcal{I}$. The results obtained for the structures ${\bf
S_1}$ and ${\bf S_2}$ (Fig. \ref{compo}) suggest two types of
distributions. In Fig. \ref{compo}a (structure $\bf S_1$) the states
$\mathcal{E}$ and $\mathcal{I}$ are highly correlated whereas in
Fig. \ref{compo}b (structure $\bf S_2$) $\mathcal{E}$ and
$\mathcal{I}$ seem to be uncorrelated with each other.

Quantitatively, we measure the correlation between the
two structures by computing i) $\chi_m$, the correlation coefficient
(also called the Pearson's correlation coefficient) for the average
number of native contacts $m(n)$ and ii) $\chi_{\delta}$ the
correlation coefficient for the
variation of $m(n)$ along the chain, $\delta(n)=m(n+1)-m(n)$. These are defined by:
\ben
\nonumber
\chi_m &=&\frac{N\displaystyle\sum m_i(n)m_e(n)-\displaystyle\sum
  m_i(n)
\displaystyle\sum
  m_e(n)}
{\displaystyle\prod_{k=i,e}\sqrt{N \displaystyle\sum  m_k(n)  m_k(n)-
\displaystyle\sum m_k(n)\displaystyle\sum
  m_k(n)}},\\
\nonumber
\chi_\delta &=&\frac{N\displaystyle\sum \delta_i(n)
  \delta_e(n)-\displaystyle\sum
  \delta_i(n)
\displaystyle\sum
  \delta_e(n)}
{\displaystyle\prod_{k=i,e}\sqrt{N \displaystyle\sum
    \delta_k(n)  \delta_k(n)-
\displaystyle\sum \delta_k(n)\displaystyle\sum
 \delta_k(n)}}.
\een
The sub-indexes $i$ and $e$ refer to the states that we are comparing, i.e. $\mathcal{I}$ and
$\mathcal{E}$. 
The sums in $\chi_m$ run over all the monomers $n=1,..,N$ whereas the
sums in $\chi_\delta$   run over all the $N-1$ first
monomers. Values of correlation coefficients close to $0$ reflect a low
correlation between the structures whereas values close
to $1$ reveal large correlations. Negative values indicate anticorrelation.

We have then investigated, as exhaustively as possible, the
native state properties that lead to a similarity between
$\mathcal{I}$ and $\mathcal{E}$. 
In Fig. \ref{examples}, we report four examples of such structures that show
a three-state behavior and that have different degrees of similarity 
between $\mathcal{I}$ and $\mathcal{E}$.

Why we expect the states $\mathcal{I}$ and $\mathcal{E}$ to differ?
First of all, during the folding at zero force the monomers tend to
form native contacts independently of their position along the polymer
chain. In contrast, the intermediate state with force has the
core/extended-chain structure described above that is energetically
favored due to the stretching effect of the force. Despite of this
difference, most of the structures we investigated show a strong
correlation ($\chi_\delta$ biased towards $1$) for the variation
$\delta(n)$ of the average number of native contacts (see
Figs. \ref{compo}, \ref{examples}).

Next, we have observed that monomers in $\mathcal{I}$ tend to locally
  form crankshafts (see Figs. \ref{3state} and \ref{examples}. see
  also the illustration shown at the beginning of Section I for the
  shape of a crankshaft). Moreover, structures with a high similarity
  between $\mathcal{I}$ and $\mathcal{E}$ also show a fairly high
  content of monomers that form crankshafts in $\mathcal{N}$ (see for
  instance the structures ${\bf S_2}$ in Fig. \ref{structures} and the
  structures in Fig. \ref{examples}a, Fig. \ref{examples}b and
  Fig. \ref{examples}d). A crankshaft arrangement of monomers reflects
  the formation of non-covalent bonds between monomers that are close
  to each other along the polymer chain.  From the point of view of
  real proteins, this would suggest that the interaction of sub-units
  that are close to each other along the amino acid chain is necessary
  for $\mathcal{I}$ and $\mathcal{E}$ to be similar.

\paragraph*{{\bf On and off-pathway states. }}

The extension trace of Fig. \ref{3state} (see also Fig. S5 and Fig. S6
in Supp. Mat.) shows that the last folding step starts from
$\mathcal{I}$. A similar observation has led Cecconi {\it et. al} to
argue that $\mathcal{I}$ is on-pathway. However, we cannot discard the
possibility of the presence of additional intermediate states
off-pathway having the same molecular extension, i.e. misfolded states
(Fig. \ref{4state}).  We then propose the following experimental force
jump protocol to detect and quantitatively measure the fraction of
misfolded states.  Each time the system folds into $\mathcal{I}$, we
relax the force to zero and compute the distribution of folding
times. In the presence of misfolded states, one should get a
bimodal distribution composed of a short-time contribution corresponding
to on-pathway states and a long-time tail corresponding to off-pathway
states. Indeed, misfolded states are expected to be separated from
$\mathcal{N}$ by high energy barriers that slow down the folding
dynamics leading to large folding times \cite{gutin,onuchic}.

We have carried out numerical simulations of this force jump protocol
(i.e. we relax the force to zero once the system has a number of
contacts corresponding to $\mathcal{I}$) in two cases: when only
on-pathway intermediate states are present and when a mix of
on-pathway and off-pathway states are present. In most cases we studied we found that 
$\mathcal{I}$ was on-pathway. A convenient way to generate misfolded states
is to consider a structure showing only on-pathway states and then add solvent conditions 
that favor the formation of misfolded states. We
include the effect of hydrophobic interactions between the amino
acid side chains of a protein and the water molecules in solution by
introducing an additional energy term $e_h$ for each interaction between
a molecule of the solvent (corresponding to a free node on the
lattice) and a monomer. The overall energy contribution for a monomer is 
$pe_h$ where $p$ is the number of nearest neighbor free nodes of that
  monomer. 
$e_h>0$ favors
hydrophobicity by increasing the interactions between the monomers.  For
the sake of simplicity, we have taken a single value of $e_h$ for all
monomers. However, one could do more general and introduce a value of
$e_h$ for each individual monomer by adding specific (positive or
negative) contributions to control the degree of hydrophobicity of each
monomer. The latter procedure has been used to model the effect of a
denaturant on the folding transition \cite{pande}.

\begin{figure}[t]
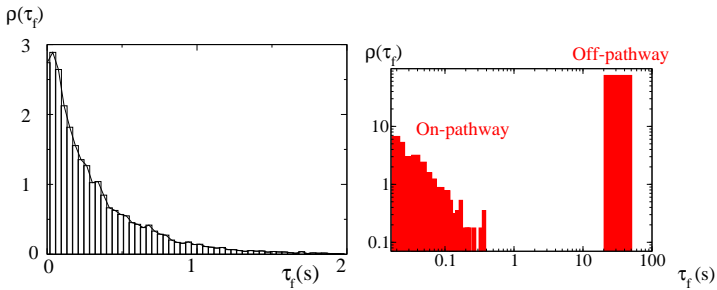

\begin{center}
\centerline{\includegraphics[scale=0.19]{fig_extra_5.eps}\includegraphics[width=0.26\textwidth]{fig7.eps}}
\caption{Left panel: Distribution of folding time $\tau_f$ after
setting the initial force to zero once $\mathcal{I}$ is reached in
the absence of
hydrophobic effects for the  structure ${\bf S_1}$ with $\Delta=1.7 \,
k_B T^*$ and $f=13.2$ pN. In this case, we observe 
only on-pathway states. Right panel:
Same distribution of folding time $\tau_f$ but in the presence of
hydrophobic effects that lead to off-pathway intermediate states.
The values are $\Delta =1.67 \, k_B T^*$, $f=13.2$ pN and $e_h=0.5 \,
k_B T$. The rightmost vertical bar counts for trajectories that have
$\tau_f>20$ s. Because the size of the systems we run our simulations 
is small, when $\mathcal{I}$ is reached we constrain the system to
keep a number of native contacts larger than those in
$\mathcal{I}$. In this way we reduce finite-size effects and obtain a clear separation of
timescale between on-pathway and off-pathway states.
\label{misfoldistri}}
\end{center}
\end{figure}

Fig. \ref{misfoldistri} reports the distribution of folding times for
${\bf S_1}$ that shows the presence of misfolded states.  Without
hydrophobicity ($e_h=0$), the temperature and force values used are such
that there are only on-pathway states, thus leading to a smooth
monotonic distribution of folding times.  By adding
hydrophobicity, i.e. $e_h>0$, monomers tend to interact more with each
other.  Although the temporal evolution of $r_{\mbox{\scriptsize end}}$
is similar to that observed in the $e_h=0$ case, we actually obtain a
mixture of on-pathway and off-pathway states.  The former contributes to
the short time distribution of Fig. \ref{misfoldistri} whereas the
latter corresponds to the contribution at very large times. By
separately integrating out each part of the distribution, we are able to
measure the fraction of on/off-pathway trajectories that lead to on/off
pathway states respectively.  For the example shown in
Fig. \ref{misfoldistri}, we get $42 \% \pm 5 \%$ and $58 \% \pm 5 \%$ of
on/off-pathway folding trajectories respectively. Force jump protocols could be
implemented in optical tweezers and AFM experiments to measure the fraction of on/off-pathway folding trajectories.

\section{Experimental protocols and the intermediate state}
\label{core}

Determining the structure of non-native states of nucleic acids and
proteins remains a major experimental challenge in modern biophysics.
For instance, even the structure of the denatured state of the lysozyme
protein that has been studied over half a decade is still unresolved
\cite{fitzkee}.  In this regard, the use of single-molecule techniques
appears as a promising tool to identify kinetic pathways and
intermediate states \cite{ritort}.  In this section, we propose specific
protocols in single molecule pulling experiments aiming to determine,
within {\it one} amino acid accuracy, the location of the core in
proteins with an intermediate state. Implementation of these protocols require 
well known methodologies in protein biophysics.

A useful method to determine the structure of $\mathcal{I}$ consists in
measuring the unfolding/folding kinetic rates, $k_u$ and $k_f$,
associated to the transition $\mathcal{I} \leftrightarrow \mathcal{S}$
after modifying the protein in various ways. These rates are obtained by
recording, at a given force, the molecular extension of the protein and
measuring the inverse of the average residence time of the protein in
each state $\mathcal{I}$ and $\mathcal{S}$ (see Fig. \ref{3state}).

\begin{figure}[t]
\begin{center}
\centerline{\includegraphics*[width=0.18\textwidth]{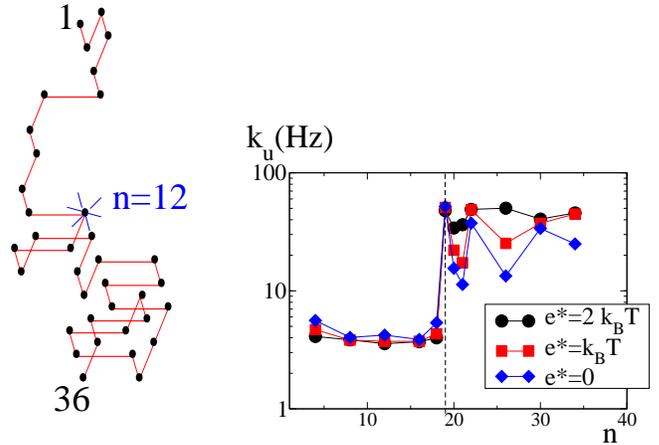}
\includegraphics[width=0.3\textwidth]{fig8b.eps}}
\caption{Point mutation protocol. Left: Unstructured intermediate
configuration. The star indicates a mutated amino acid. Right: $k_u$ a
function of the location $n$ of the mutation along the chain for
different values of the hydrophobicity of the mutated monomer,
$e^*$. The structure is ${\bf S_2}$, $\Delta=1.2\, k_B T^*$ and $f=9$
pN. The dashed line indicates the position of the first monomer
($n_f=19$) inside the core. The core is composed of all the monomers
that follow up from that monomer until the end of the chain.
\label{mutation}}
\end{center}
\end{figure}

\paragraph*{{\bf A "$\phi$-value" force protocol.}}

A possible modification of the protein consists in selectively mutating
an individual amino acid.  The idea of this method is reminiscent of the
$\phi$-value technique used in bulk measurements \cite{fersht2}.  In our
case, we consider a heteropolymer where initially $e_h^i=0$ for all
$i$. We then select one monomer $i$ and assign new values for the
interaction energies $E_{ij}$ between that monomer and the other
monomers $j$ of the chain. We also increase the degree of hydrophobicity
of that monomer $i$ by setting $e_h^i=e^*>0$ while keeping the rest of
the $e_h^i$'s equal to zero.  In Fig. \ref{mutation} we report for ${\bf
S_2}$ the values of $k_u$ as a function of the location $n$ of the mutation along the
chain and for different values of $e^*$. One can clearly
see a transition separating low and high rates that is distinctly
located at the edge of the core.  Low rates correspond to mutations on
the free chain whereas high rates correspond to mutations inside the
core.  The larger $e_h$, the sharper the transition, which suggests the
use of very hydrophilic amino acids, such as serine or threonine as
point mutations.

\begin{figure}[t]
\begin{center}
\centerline{\includegraphics*[width=0.18\textwidth]{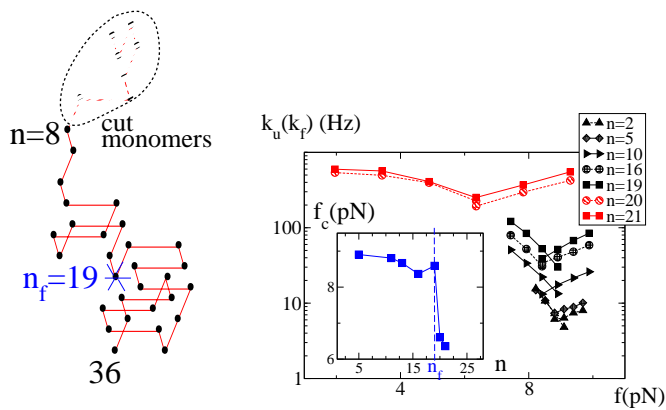}
\includegraphics[width=0.3\textwidth]{fig9b.eps}}
\caption{Cutting protocol. Left: Unstructured intermediate
configuration. The black dashed circle indicates the cut in the
chain. Cutting at the monomer $n$ along the heteropolymer
means ``removing the monomers 1, 2, .., n-1''. Right: We report the values of the ``unfolding'' and
``folding'' rates for the reaction $\mathcal{I} \leftrightarrow
\mathcal{S}$ as a function of the applied force $f$, which leads to the so-called
chevron plots. The different
chevron plots correspond to different locations of the cutting of the
heteropolymer. The structure is  ${\bf S_2}$ with $\Delta=1.2 \, k_B
T^*$. Insets: Critical
forces as a function of the cutting position $n$.  The vertical dashed line marks the
position of the first monomer ($n_f=19$) that separates the core from
the extended chain. We find that, at $n_f$, the value of the critical force (where unfolding and folding rates 
are equal) suddenly drops (compare the chevron plots for $n=19$(black) and $n=20$(red)).
\label{chevron}}
\end{center}
\end{figure}

From an experimental point of view, a single mutation may not be
sufficient to distinguish a transition because of the too small
differences in the rates. We then suggest a multiple-points mutation
analysis: instead of mutating a single amino acid, two or more
successive amino acids can be mutated. This helps to identify more
clearly the transition but also leads to a less precise location of the
position of the edge of the core (Fig. S7 in Supp. Mat.).

\paragraph*{{\bf Cutting the proteins.}}
According to Kramers-Bell theory, the transition rates between
$\mathcal{I}$ and $\mathcal{S}$ depend exponentially on the applied
mechanical force, showing a chevron-like shape \cite{finkelstein} -- see
Fig. \ref{chevron}. In these kind of plots we represent the unfolding ($k_u$) and folding ($k_f$) rates as 
a function of force. Therefore, the increasing (respectively decreasing) curves correspond
to the dependence of the rates $k_u$ (respectively $k_f$) of the
transition $\mathcal{I} \to \mathcal{S}$ (respectively $\mathcal{S} \to
\mathcal{I}$). The crossing 
point of a chevron plot ($k_u=k_f$) is located at the value of the force where
both rates ($\mathcal{I} \to \mathcal{S}$ and $\mathcal{S} \to
\mathcal{I}$) are identical. This is the critical force where the two species
($\mathcal{I}$ and $\mathcal{S}$) are equally probable.  We have measured these rates in $\bf S_1$
and ${\bf S_2}$ after cutting off the extremities of the chains at certain
locations, i.e. leading to a shorter polypeptide chain.
Fig. \ref{chevron} shows the chevron plots for ${\bf S_2}$ as we keep the
extremity fixed at one end of the core and progressively reduce the
length of the chain.  We see a sharp transition, characterized by a drop
of the critical force (insets of Fig. \ref{chevron}), when the cut is
done inside the core. In this case, $\mathcal{I}$ looses its stability
because of the spoiling of the core, a rather intuitive result.  This
allows again to locate the core with one monomer accuracy. Other similar
modifications where protein interactions are changed are shown in
Fig. S8 in the Supp. Mat..

\paragraph*{{\bf Circular permutations.}}

\begin{figure}[t]
\begin{center}
\centerline{\includegraphics*[width=0.18\textwidth]{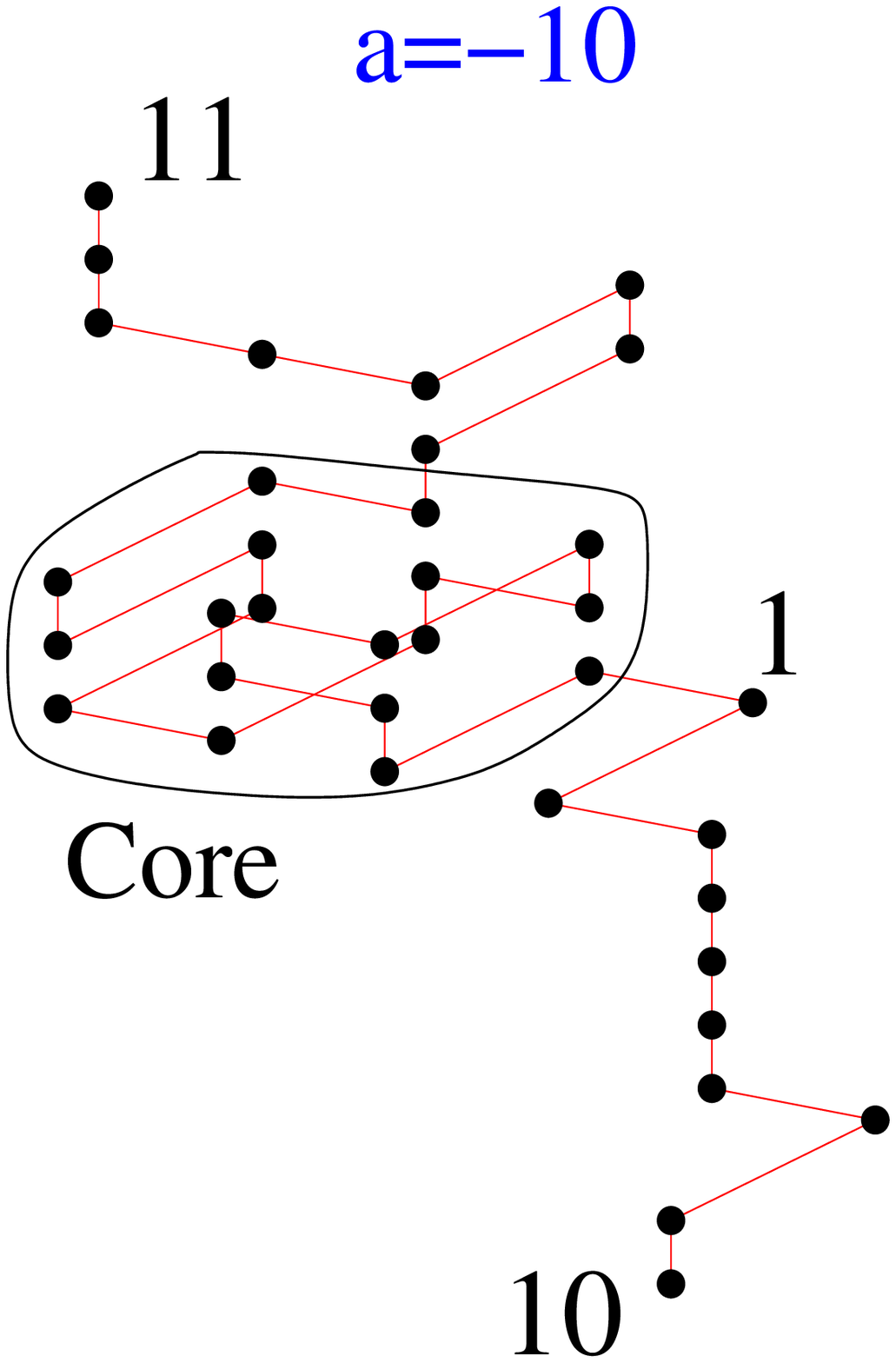} \hfill
\includegraphics[width=0.3\textwidth]{fig10b.eps}}
\caption{Permutation protocol. Left:  Unstructured
intermediate configuration after circularly permuting ${\bf S_2}$ by
$a=-10$. The numbers indicate the initial position of the monomers in
${\bf S_2}$.  Right: Chevron plots for circular permutations of ${\bf
S_2}$ by $a$. $\Delta=1.2 \, k_B T^*$.  Inset: Critical forces as a
function of $a$.  The vertical dashed lines indicate $a$ values where the core 
dissociates.
\label{perm}}
\end{center}
\end{figure}

Circular permutations are useful modifications that allow to investigate
the stability of native structures.  In this case, new polypeptide
chains are obtained by shifting all the amino acids in the original
chain by a certain amount $a$. An amino acid at the position $i$ will
then go to the position $i+a$ (modulo the number of amino acids in the
protein) where $a$ can be positive or negative.  We have measured the
rates $k_u$ and $k_f$ in heteropolymers obtained by circular
permutations of ${\bf S_2}$.  We find, again, transitions when the
circular permutation dissociates the core. The two transitions are found
at both edges of the core (right and left, see Fig. \ref{perm}) and are
characterized by a sudden drop in the critical force.

We have finally investigated an experimental protocol in which we change
the location of the applied force along the chain. In this case, the
presence of undesired interactions involving the monomers that are not
pulled by the force makes difficult the analysis of the traces.  The
traces are indeed noisy due to the formation of new states that are very
unlikely when stretching from the very ends of the heteropolymer.
The important problem about how mechanical unfolding depends on
  the location of the force entails a more detailed study that we do
  not pursue here.

\section{Conclusion}

In many respects designed on-lattice heteropolymers are crude
approximations to real proteins. Yet, it seems again, that these
models share common features with the folding of single-domain
proteins. In particular these models seem appropriate to investigate
the three-state behavior that has been recently observed in
single-protein force experiments \cite{cecconi}.  What is the
link between the present lattice model results and real wild-type
proteins? It is important to make clear the limitations of the current
approach. Although most of our study has been inspired by recent
results in RNaseH it remains a challenge to establish a clear
connection between the native three dimensional structure of a real
protein and the topology of the native structure used in heteropolymer
models. Let us stress that lattice models are phenomenological
models useful to design specific free energy landscapes capable of reproducing
different kinetic scenarios for folding (e.g two-states, three-states,
intermediate states on/off pathway, correlated/uncorrelated early and 
intermediate states and so on). From this perspective we expect that the
phenomenology described here is quite general and probably observed in
proteins other than RNase H.

Interestingly, the stabilization by mechanical force of a unique
intermediate state suggests possible ways to experimentally infer its
structure. We have found that this state is composed of an unstructured
and stretched part of the polypeptide chain plus a rigid core that
corresponds to some part of the native state. This result might be
specific to the details of the model, yet the competition between different
types of low entropy regions along the polypeptide chain (a compact core
versus an extended chain) could be reasonably argued to be the generic
driving force for the formation of unstructured extended chains. It must
be emphasized, however, that our model does not include side chain
interactions. These are known to induce a large entropy loss upon
folding due to the excluded volume interactions present in the packed
native state \cite{BroDil94,KliThi98}. Therefore we cannot exclude a
scenario where the large entropy of the side chains might induce a
molten globule like intermediate state in force where the protein keeps
a single native-like core with freely moving side chains\cite{ShaFin89}.
Our simulations also reveal (see Fig. S10) that the
presence of a rigid core is not necessarily correlated with the
hydrophobicity of the monomers in the chain. This suggests that,
although amino acid composition can facilitate the formation of a core,
an excess of hydrophobic monomers is not a necessary requirement for its
formation.

Although real proteins are too complex to be modeled with "beads and
sticks in regular lattices", these models are useful to infer possible
experimental protocols to probe the intermediate state.  The
experimental protocols we propose in this work (point mutations, cutting
the polypeptide chain and circular permutations) are well known in
protein biophysics and could be used to distinguish between a molten
globule and an unstructured extended state.  Indeed, if these
modifications of the protein lead to the same types of transitions as in
Fig. \ref{mutation}, \ref{chevron}, \ref{perm}, it is likely that the
intermediate state is composed of a core plus an unstructured extended
chain.  In the case of a uniform dependence of the rates this would
suggest that the intermediate state resembles more a molten-globule
structure where no rigid core is present. More generally, these
techniques could be applied to precisely determine the location of the
disordered and ordered domains in unstructured proteins.

Finally we have shown that force measurements can also be used to
highlight the presence of misfolded states, and to quantify the relative
fraction of on/off-pathway trajectories.  From this perspective force
measurements suggest the possibility of probing the shape of the free
energy landscape in proteins and investigating the glassy behavior of
proteins at low temperatures \cite{bryngelson,hyethi,brujic}.

\begin{acknowledgments}
We are grateful to C. Bustamante, C. Cecconi, M. Manosas and S. Marqusee for useful suggestions.
I. J acknowledges financial support form the European network STIPCO,
Grant No. HPRNCT200200319. F. R acknowledges financial support from
the Ministerio de Eduaci\'on y Ciencia (Grant FIS2004-3454 and
NAN2004-09348) and the Catalan government (Grant SGR05-00688).
\end{acknowledgments}

\maketitle

\end{document}